# Nanodiamond–quantum sensors reveal temperature variation associated to hippocampal neurons firing

G. Petrini[1,2], G. Tomagra[3], E. Bernardi[1], E. Moreva[1], P. Traina[1], A. Marcantoni[3], F. Picollo[2,4], K. Kvakova[5,6], P. Cigler[5], I.P. Degiovanni[1,4], V. Carabelli[3], M. Genovese[1,4,*]

[1]Istituto Nazionale di Ricerca Metrologica, Strada delle cacce 91, Turin, Italy

[2]Physics Department and NIS Inter-departmental Centre, Torino, Italy

[3]Department of Drug and Science Technology - University of Torino, Corso Raffaello 30 Torino, Italy and NIS Inter-departmental Centre, Torino, Italy

[4]Istituto Nazionale di Fisica Nucleare (INFN) Sez. Torino, Torino, Italy

[5] Institute of Organic Chemistry and Biochemistry of the Czech Academy of Sciences, Flemingovo nam. 2, 166 10 Prague 6, Czechia

[6]Institute of Medical Biochemistry and Laboratory Diagnostics, First Faculty of Medicine, Charles University, Katerinska 1660/32, 121 08 Prague 2, Czechia

*Corresponding author. Email: m.genovese@inrim.it

## Abstract

Temperature is one of the most relevant parameters for the regulation of intracellular processes. Measuring localized subcellular temperature gradients is fundamental for a deeper understanding of cell function, such as the genesis of action potentials, and cell metabolism.

Here, we detect for the first time temperature variations (1°C) associated with potentiation and depletion of neuronal firing, exploiting a nanoscale thermometer based on optically detected magnetic resonance in nanodiamonds. Our results provide a tool for assessing neuronal spiking activity under physiological and pathological conditions and, conjugated with the high sensitivity of this technique (in perspective sensitive to < 0.1°C variations), pave the way to a systematic study of the generation of localized temperature gradients.

Furthermore, they prompt further studies explaining in detail the physiological mechanism originating this effect.

**1. Introduction**

On the one side temperature regulates the speed of ion channel opening [1], the pattern activity of a firing neuron [2], the vesicular dynamics at the presynaptic terminal [3], on the other side intracellular temperature is affected by a variety of biochemical reactions occurring during cell activity. Pioneering findings dating back to the late 70's [4,5] associated temperature increases to the impulse propagation in non-myelinated fibers of the olfactory nerve and, more recently, a theoretical explanation about the heat production and absorption by neurons during nervous conduction has been formulated [6]. Intracellular temperature variations have been probed to detect the phases of cell-cycle division [7] and the mitochondrial activity [8]. In the brain, temperature fluctuations are likely associated with changes in neuronal functions such as the genesis of action potentials, secretion at the level of synaptic terminals, or transmembrane ion transports [9–11]. Variation of intracellular temperature is also related to changes in cell metabolism, as observed through positron emission tomography (PET) and functional magnetic resonance imaging (fMRI) [12].

Besides its fundamental role in cell physiology, the temperature may be altered under pathological conditions, such as in cancerous cells [13], which display higher metabolic activity, in neurodegeneration, such as Parkinson's or Alzheimer's diseases, where the process of abnormal protein aggregation is temperature-dependent [14], or during the onset of *malignant hyperthermia*, a pathology that causes rapid temperature increases following excessive muscle contraction [15].

Furthermore, an increase in local temperature, up to a few Celsius degrees, has been detected as triggered by laser heating [16], calcium stress [17], direct electric stimulation [18] or using drugs that increase the heat produced during cellular respiration (17,19,20). These results, in particular the ones related to the highest temperature variations, stimulated a debate since the power needed to justify this temperature growth has been calculated to exceed a few orders of magnitude what is expected by thermodynamic considerations in a model where thermal diffusion is characterized by conductive regime [20]. Nonetheless, other authors [21,22] suggested that, by taking into account the inhomogeneity of the cells, this gap can be strongly reduced.

However, detection of temperature fluctuations at the subcellular level still represents an ongoing challenge and several intracellular thermometry techniques are reported in the literature, ranging from fluorescent molecular thermometers [23], quantum dots [24] or rare-earth nanoparticles [25]. Compared to the above-mentioned methods, nanodiamonds (ND) show a better biocompatibility [26,27], insensitivity to biological environment [28], more stable photoluminescence (PL) and a lower noise floor [29].

All-optical temperature measurement methods, based on the temperature-dependent shift of the luminescence spectra of color centers in diamond such as silicon-vacancy (SiV), germanium vacancy (GeV), and tin vacancy (SnV) with strong zero-phonon line (ZPL), can be used [30]. These methods are promising, but for the moment, apart from one very invasive experiment [31], there is no reported biosensing application and currently the most promising approach exploits nitrogen-vacancy (NV) centers [29].

For these reasons, NV color centers in ND combined with Optically Detected Magnetic Resonance (ODMR) technique can assume a dominating position in thermometry for biological applications [32]. Nevertheless, ODMR measurements protocols are based on microwave irradiation of the samples, which makes this technique more elaborated and requires awareness on the exploitable optical and microwave power providing the required sensitivity without damaging the cells.

ND was used for the first time *in vitro* [33] measurements (of a heated gold nanoparticle inside a cell), subsequently an *in vivo* experiment was performed inside *Caenorhabditis elegans* worms, reaching a sensitivity as low as 1.4 °C/Hz $^{-1/2}$. In the same experiment, thermogenic responses have been monitored during the chemical stimuli of mitochondrial uncouplers [16]. Temperature gradients have been mapped at the subcellular level into a single human embryonic fibroblast [7]; intracellular temperature mapping has been performed also in cultured primary cortical neurons, employing microelectrode arrays (MEAs) to demonstrate that the presence of NDs in primary cortical neurons does not elicit a neurotoxic response [34], in good agreement with our previous findings [35]. Finally, heterogeneous temperature variations have been coupled with $Ca^{2+}$ increases in HeLa cells [36].

Here, for the first time, we quantify the direct correlation of intracellular temperature variations with the modulation of neuronal activity. We demonstrate that sensors based on NV centers in NDs, interrogated via ODMR techniques, reveal up to 1°C temperature variation when the spontaneous firing of hippocampal neurons is potentiated by Picrotoxin or the 0.5°C temperature decrease when the neuronal activity is silenced by a solution containing tetrodotoxin and cadmium chloride.

## 2. Results

**Figure 1a** shows a simplified scheme of the apparatus used for the experiment. Temperature variations were detected using an ODMR measurement protocol, consisting of simultaneously applying a variable frequency microwave field and non-resonant laser radiation to the sample while acquiring the PL intensity (photon counts) emitted by a single ND (Figure 1b).

By applying this protocol, a dip in the collected PL intensity is observed when the microwave frequency matches the exact resonant frequency $D_{gs}$ between the $|m_s=0\rangle$ and $|m_s=\pm 1\rangle$ electronic spin states of the NV center. Since the value of $D_{gs}$ depends on temperature, this technique allows determining the temperature at the ND location by measuring a frequency shift.

The temperature variation ΔT measured with the ND sensor is estimated through a direct PL change observation according to the equation:

$$\Delta T = \frac{\Delta \tilde{F}}{slope \cdot \partial D_{gs}/\partial T} \quad (1)$$

where $\Delta \tilde{F}$ is the actual measured physical signal (difference in the photocounting rate), *slope* represents the constant of proportionality that connects $\Delta \tilde{F}$ to the resonance frequency shift $\Delta D_{gs}$. Finally, $\partial D_{gs}/\partial T$ represents the coupling constant. Under ambient conditions the coupling constant $\partial D_{gs}/\partial T$ is estimated as -75 kHz/°C for a bulk diamond (see Materials and Methods).

The acquisition time of all PL measurements was 60s. This value was chosen as a trade-off between the fast measurement and the precise temperature estimation. We underline that our choice guarantees cell viability and, in principle, allows for time-resolve measurements on the scale of metabolic processes.

To exploit the ODMR protocol for measuring intracellular temperature variations, hippocampal neurons (10 days in vitro) were incubated for 5 hours with 0.6 µg/ml NDs (average diameter 185 nm, PDI 0.115), which is far below the cytotoxicity threshold (> 250 µg/ml). ND internalization (Figure 1c) was assessed by standard confocal imaging (see Materials and Methods).

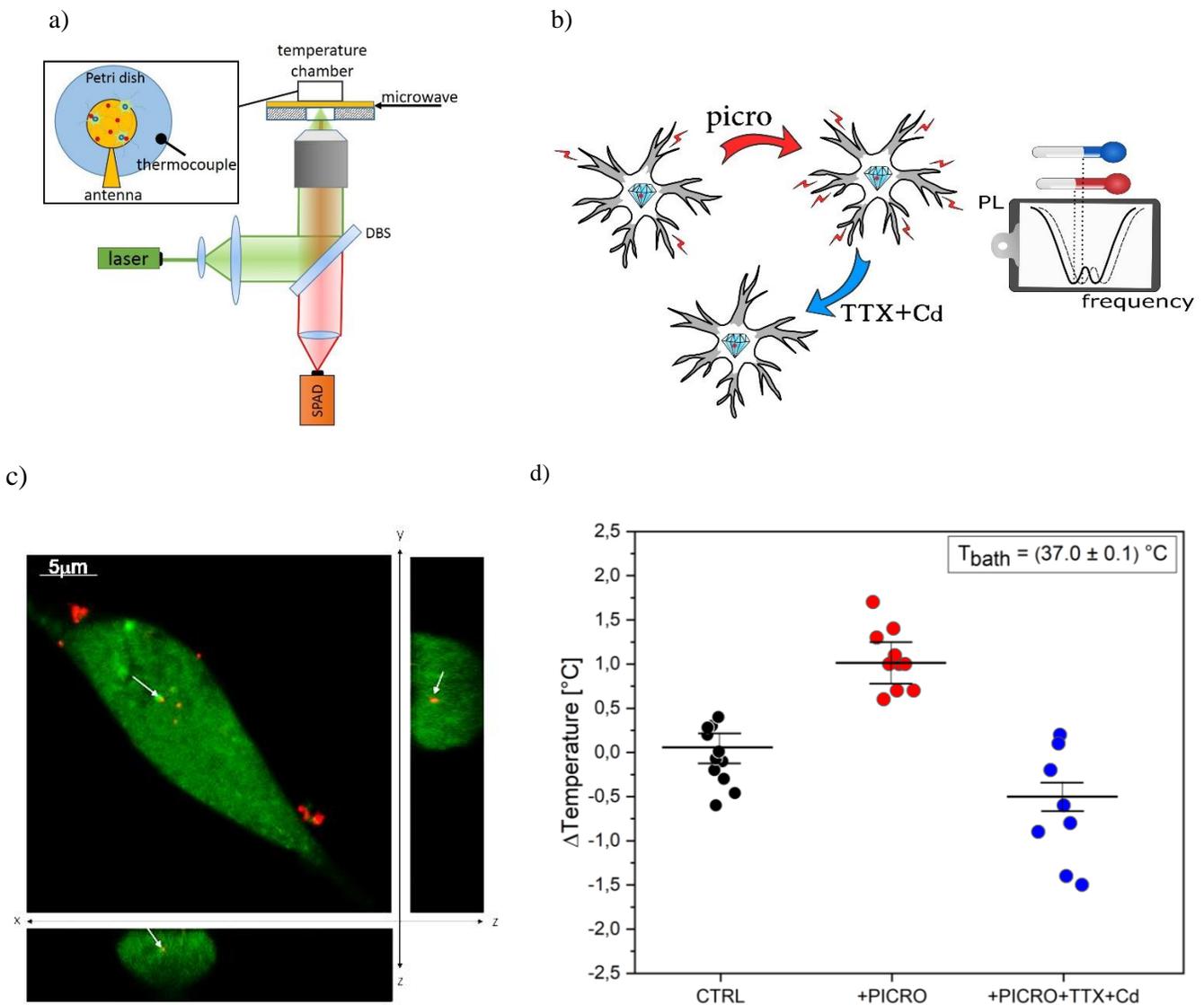

**Figure 1.** Illustration of the experiment. a) Simplified scheme of single-photon confocal ODMR setup b) The ODMR measurements are performed under control conditions (CTRL), after stimulation with Picrotoxin and after the addition of TTX+Cd. The frequency shift in the ODMR spectrum (dashed line) is associated with the temperature variation recorded by the ND sensor, c) Confocal fluorescence micrograph of hippocampal neurons incubated with 0.6 µg/ml ND for 5 hours. The cytoplasm is stained in green, the red emission is from NDs. The entire field and cross-sections (XZ and YZ) are shown. White arrows show one internalized ND. d) Boxplot of temperature variations with standard deviations in the presence of saline Tyrode solution (CTRL, black circles), after addition of Picrotoxin (PICRO, red circles), after addition of tetrodotoxin and cadmium chloride (TTX+Cd, blue circles), see text for details.

To detect the temperature variations associated with different patterns of neuronal activity, recordings were performed in three conditions: (*i*) under control conditions (CTRL), in which the hippocampal network was spontaneously firing (external Tyrode solution) [28], (*ii*) after addition of 100 µM of Picrotoxin, a selective $GABA_A$ inhibitor, which drastically potentiated the firing activity, (*iii*) after subsequent addition of 0.3 µM tetrodotoxin (TTX) and 500 µM cadmium chloride (Cd), depleting the firing activity. Bath temperature was kept at 37°C and the photocounting rate was performed from internalized NDs. With reference to the above conditions, a more detailed account of the three measurement condition follows hereby.

-Condition *(i)*: To exclude that the perfusion system could induce some temperature variations, the photocounting rate $\tilde{F}$ from an internalized ND was measured before and after perfusing the cells with Tyrode solution, respectively (CTRL, in Figure 1d). From the photoluminescence difference $\Delta\tilde{F}$, the temperature variation was estimated according to Equation 1. In these conditions we revealed an average temperature variation ΔT = (0.05±0.15) °C (weighted average evaluated on a sample with a numerosity N=11), proving that no significant temperature variations are associated with thermal exchanges of the perfusion system.

After this preliminary test, the next step was to assess whether changes in neuronal firing could be associated to temperature variations.

-Condition *(ii)*: ΔT was estimated by comparing $\tilde{F}$ before and after perfusing the cells with Picrotoxin solution. The addition of the $GABA_A$ inhibitor caused a significant temperature increase, ΔT = (1.02 ± 0.24) °C (N=10), associated with the increased firing rate (Figure 2a). As a more quantitative corroboration of our conclusions, a thorough statistical uncertainty analysis was performed (see Materials and Methods) to confirm that a significant localized temperature variation occurs in a neuron deriving from sustained firing activity ($1.12 \cdot 10^{-8}$ significance in a *Welch t*-student test).

-Condition *(iii)*: network activity was silenced by means of the Tyrode solution enriched with TTX+Cd. In this case, ΔT was measured by comparing $\tilde{F}$ before and after perfusing the cells with Picrotoxin+ TTX+Cd solution. Under this last condition, a significant temperature decrease ΔT = (-0.50 ± 0.17)°C (N=8) was revealed with a $2.21 \cdot 10^{-6}$ significance in a *Welch t*-student test (see Materials and Methods). These data demonstrate that both potentiation and silencing of the neuronal network activity could be assessed by temperature variations and that the hippocampal network, exhibiting basal spontaneous activity, displays higher temperatures that the completely silenced network.

Another set of experiments has been carried out for measuring the average temperature variation associated to non-internalized ND before and after perfusing the cells with Picrotoxin solution: in this case no significant difference with respect to the control conditions: $\Delta T$ = (-0.04 ± 0.23) °C (N = 7) was found. This crossover trial excluded that temperature variation can be due to other external factors than the drug-potentiated neuronal activity.

Furthermore, it has been tested whether the laser exposure, microwave radiation and incubation with NDs could affect the spontaneous firing activity of hippocampal neurons during ODMR measurement protocol (as detailed in Methods). As shown in Figure 2, both the spontaneous firing rate and the action potential waveform were not significantly altered by laser and ND exposure, confirming that cell excitability and ion channel functioning are preserved, in good agreement with our previous findings [26].

Finally, in parallel with the thermometry experiment, the effect of $GABA_A$ receptor blockade on burst-firing of hippocampal neurons was assessed by comparing the spontaneous burst ratio of cells responses before and after application of Picrotoxin. These recordings have been performed by means of conventional MEAs (MCS, multichannel system). An expected significant increase of the spontaneous activity [37] (from 2.1 ± 0.3 Hz to 3.4 ± 0.3 Hz), which is associated to disinhibition of $GABA_A$ receptors (Figure 2a), and a complete silencing of the network in the presence of the Na$^+$ and Ca$^{2+}$ voltage-dependent channel blockers TTX and Cd was observed. The data presented allowed to conclude that the increase in temperature recorded in the experiment is directly related to the altered firing activity of hippocampal neurons.

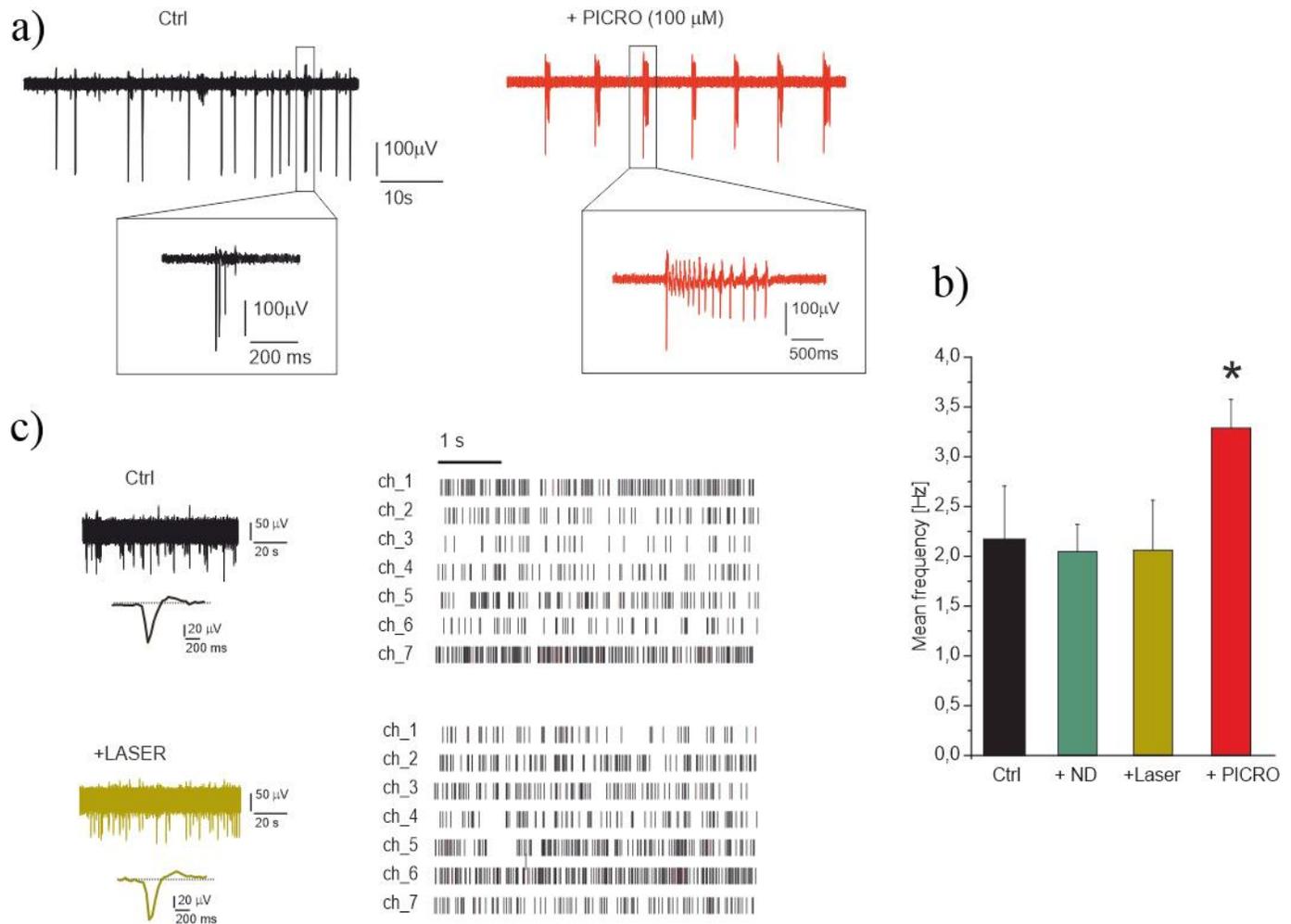

**Figure 2.** Laser exposure does not affect the hippocampal neurons. a) modulation of the firing activity by PICRO (MEA recordings): representative traces from the same electrode under control condition, + PICRO. Insets: higher magnification of single spikes and bursts. b) Histogram of mean frequency in the different experimental conditions. c) Left: Representative traces in control condition and after laser irradiation and (right) corresponding raster plot. In the raster plot, 7 representative channels (ch_1÷ch_7) are shown.

## 3. Discussion

Intracellular localized temperature gradients are associated with metabolic activity and with a variety of reactions and intracellular processes [38,39]. Real-time mapping of intracellular subtle temperature gradients and event-driven increased temperature represents a tool of the utmost interest for monitoring the functional activity of the cells, for identifying localized signaling under physiological conditions and for exploitation as a diagnostic tool.

Here we applied the ODMR technique to cultured hippocampal neurons for monitoring temperature variations in different conditions of network excitability. We provided evidence that a significant temperature increase can be correlated with the altered firing activity of cultured hippocampal neurons, in our case induced by Picrotoxin, and that this phenomenon can be successfully observed at the single-cell level exploiting nanosensor based on NDs capable of being in perspective sensitive to <0.1°C variations. We underline that these nanosensors, carefully prepared and selected, are able to provide a fast temperature measurement with extraordinary spatial resolution, in perspective even below the diffraction limit [40].

The action potential generation and propagation is not the unique energy-requiring process involved in neuronal activity, as this may also involve maintenance of the membrane resting potential, neurotransmitter release and uptake, vesicular recycling and presynaptic $Ca^{2+}$ currents [39]. The firing activity is correlated with the total energy consumed by neuronal activity. This has been calculated from anatomic and physiological data [38] and experimentally verified using fMRI techniques [41]. Thus, our results prompts further studies to assess whether the observed temperature increase during perfusion with Picrotoxin is ascribed to sustained firing activity and/or potentiated cell metabolism (44).

By means of confocal microscopy observation, we demonstrate that 5 hours exposure to ND is enough to promote ND internalization and that action potential waveform remains unaffected after laser and microwave irradiation, confirming that the applied protocol for sensing temperature preserves cell excitability and ion channel functioning[26] (with perspectives for "in vivo" studies).

The measured temperature increases can be used to reveal the onset of different intracellular processes other than membrane-delimited pathways, likely involving an altered cell metabolism as here was demonstrated with Picrotoxin. Thus, our future goals will be devoted to functionalizing NDs to subcellular compartments, to detect thermogenesis at specific subcellular sites.

These results, backed by a thorough analysis of other possible alternative causes of temperature variation, pave the way to a systematic study of cell activity with impacts ranging from a better insight on current unknowns associated with cells functioning, such as e.g. the aforementioned discrepancy between experimental data and thermal diffusion models, to the research on specific pathologies.

Our findings also prompt further applications where ongoing advances in microelectrode array technology (MEA), combined with quantum sensing paves the way for experiments that take advantage from the synergy of the two techniques, such as synchronized measurements of cellular activity with metabolism processes or propagation of electromagnetic signals.

## 4 Materials and Methods

### 4.1 Experimental setup

The thermometric apparatus used for this experiment (Figure 3) was based on an Olympus IX73 inverted microscope, in which optical elements to implement single-photon confocal imaging and microwave control for ODMR measurements were integrated. A CW 532-nm laser (Coherent Prometheus 100NE, noise reduced regime, $2^{nd}$ harmonics), attenuated down to 1 mW was used to excite the NV centers in the NDs. Then an acousto-optic modulator (AOM), controlled by a pulse generator (Pulse streamer, Swabian Instruments), was applied to the laser emission to have the cell irradiated only during the measurement. Finally, a 60× air microscope objective (Olympus UPLANFL, NA = 0.67) was used for both excitation and fluorescence collection. The spot size of the focused laser beam was ~ (1.2 × 1.3) $\mu m^2$. The NV photoluminescence (PL) was filtered by a 567 nm dichroic mirror, a 650 nm long-pass filter and a Notch filter centered at 532 nm to remove the residual green laser scattering, and then collected by a single photon avalanche diode (SPAD, SPCM-AQR 15, Perkin Elmer). The emission rate from a single ND was around 300 kCounts/s.

The Petri dish, containing the cell culture with NDs, was placed in a closed incubation chamber with a temperature control (Okolab Temperature Controller, temperature stability 0.1°C). The sample could be moved via a manually or software-controlled piezoelectric XYZ scanning stage allowing selecting the area of interest. The output signal from the SPAD was fed to a data acquisition board system (National Instruments, USB-6343 BNC).

To implement the ODMR measurements, a microwave source (Keysight N5172B) was used. The output signal was then amplified (Mini-Circuits, ZHL 16W 43+) to a power of 20 dBm and fed to a homemade planar broadband antenna, which provided a strong, homogeneous, electromagnetic radiation. The Petri dish was placed on the top of the antenna.

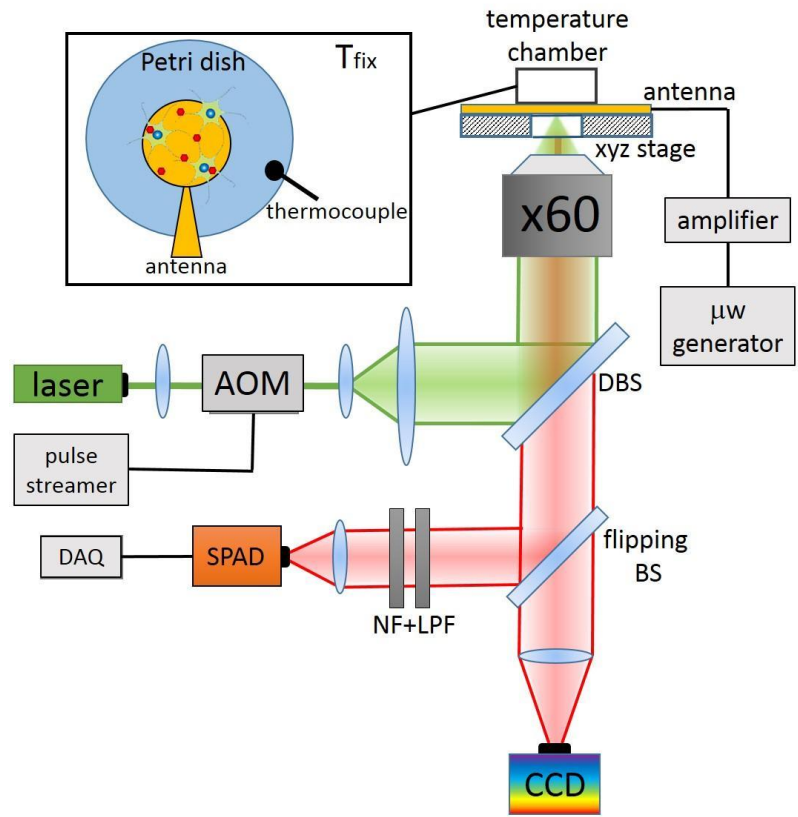

**Figure 3** Experimental setup scheme. AOM: acousto-optic modulator, DBS: dichroic beam splitter, LPF: long-pass filter, NF: notch filter, DAQ: data acquisition board. SPAD: single-photon avalanche diode. The cells are cultured on a Petri dish, placed inside an incubator (temperature chamber), which can be moved by means of a three-axis piezoelectric system (xyz stage). The temperature inside the closed incubation chamber is controlled by a PID and measured by a thermocouple.

**4.2 Experimental procedure**

The experimental procedure for detecting temperature variation using NV centers in NDs was the following.

A Petri dish containing cultured hippocampal neurons (10 DIV), previously exposed to NDs, was positioned inside the single-photon confocal microscope incubation chamber. Using the piezoelectric stage, the sample was scanned in order to select a single ND inside a cell. The ODMR spectrum was acquired for 60 s and the data was post-processed to create the differential spectrum (see "Differential ODMR spectrum experimental measurement technique"). Then, the microwave frequency was set in accordance with the minimum of the ODMR spectrum (zero of the differential spectrum). The differential signal of luminescence at this frequency was acquired for 60 s. The microwave field before and after data acquisition was always turned on to avoid possible heating transient processes.

In a successive step, solutions containing 100 μM Picrotoxin and then 300 nM TTX+200 μM Cd were added to the medium. The measurement of the differential signal was repeated at the same resonant frequency as in the previous step. Finally, the ODMR spectrum was reacquired for 400 s, recreating a new differential spectrum in order to improve the statistics and to evaluate the linear region *slope* with lower uncertainty. The temperature variation ΔT was evaluated through a direct PL change measurement according to equation:

$$\Delta \tilde{F} = slope \cdot \partial D_{gs}/\partial T \cdot \Delta T \tag{1}$$

Here $\Delta \tilde{F}$ represents the differential signal variation (difference in the photon counting rates), evaluated at the linear region of the differential spectrum and associated with a temperature variation $\Delta T$. The slope represents the constant of proportionality that connects $\Delta \tilde{F}$ to the resonance frequency shift $\Delta D_{gs}$. Finally, $\partial D_{gs}/\partial T$ represents the coupling constant.

## 4.3 Optically detected magnetic resonance (ODMR)

An important feature of the NV⁻ center is that it is possible to optically detect its spin state and optically pump it into the $|m_s=0\rangle$ sublevel. The mechanism is described as follows. Depending on the electron spin state, two different decaying paths are possible. If the system is in the state $|m_s=0\rangle$, then a spin preserving transition occurs with single-photon emission at wavelength = 637 nm (zero-phonon line, ZPL). This transition is cyclic and one detects high PL intensity when continuously illuminating the NV center.

If the electron spin state is instead $|m_s=\pm1\rangle$, two possible decay paths are possible. The first one is again spin-preserving and leads to the previously-described PL emission. The second one is non-spin-preserving and the system ends up in spin state $|m_s=0\rangle$ of the ground level (see Figure 4a). In this latter case, the system relaxes through the two metastable singlet spin states, namely $^1A$ and $^1E$. These have a transition wavelength in the infrared range, $\lambda = 1046$ nm. In this case spin changes to $|m_s=\pm1\rangle$, and thus the process is referred as "intersystem-crossing" (ISC). Due to the presence of this non-radiative decay path the spin state $|m_s=\pm1\rangle$ is less luminescent with respect to $|m_s=0\rangle$ one. Remarkably, the above-explained feature allows the optical read-out of the spin state of the NV center at room temperature. This is achieved by the method of optically detected magnetic resonance (ODMR) which consists in applying a microwave field of variable frequency, simultaneously with irradiation of a non-resonant laser able to excite the electronic state. In this case, a 532 nm pump laser is required. When the microwave frequency matches the exact resonant frequency $D_{gs}$ between the $|m_s=0\rangle$ and $|m_s=\pm1\rangle$ states, less PL is collected and therefore a fluorescence dip is observed in the ODMR spectrum (see Figure 4b). This allows the optical readout of the spin state.

It is important to underline that when NDs are used as sensors instead of bulk diamonds, the degeneration of the $|m_s=+1\rangle$ and $|m_s=-1\rangle$ levels are removed by mechanical stress. In this case, two dips are observed in the ODMR spectrum. Both of them show a shift in the same direction in the resonance frequency as the temperature varies (Figure 4d).

### 4.5 Differential ODMR spectrum experimental measurement technique

The full ODMR spectrum was acquired only once. The data was post-processed creating the differential spectrum (see Figure 4c). In correspondence to each microwave scanning frequency f, a differential photoluminescence (PL) value $\tilde{F}$ was associated, evaluated as follows: $\tilde{F} = F(f + f_{dev}) - F(f - f_{dev}))$, where $F$ represents the PL collected from the ND sensor and $f_{dev}$ is a frequency deviation chosen in order to optimize the differential graph. In our case $f_{dev} = 2\ MHz$.

The method advantage is that in correspondence of temperature variations, for a fixed frequency f the differential signal $\tilde{F}$ results linearized with respect to frequency shift around the resonant frequency $D_{gs}$:

$$\Delta \tilde{F} = slope \cdot \Delta D_{gs} \qquad (2)$$

The slope value was then obtained through a linear regression from the differential spectrum.

The differential signal $\tilde{F}$ is acquired at a single frequency chosen in the linear region of the differential spectrum. Any variation $\Delta \tilde{F}$ recorded was attributable to a shift in the resonance frequency $\Delta D_{gs}$ and therefore to a temperature variation $\Delta T$, according to Equation (1):

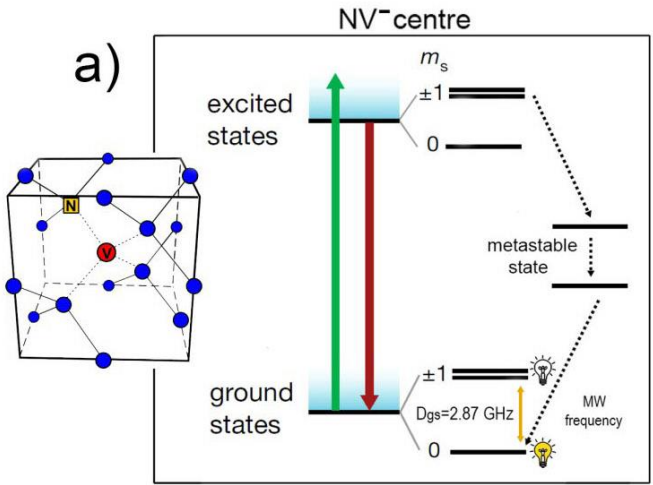
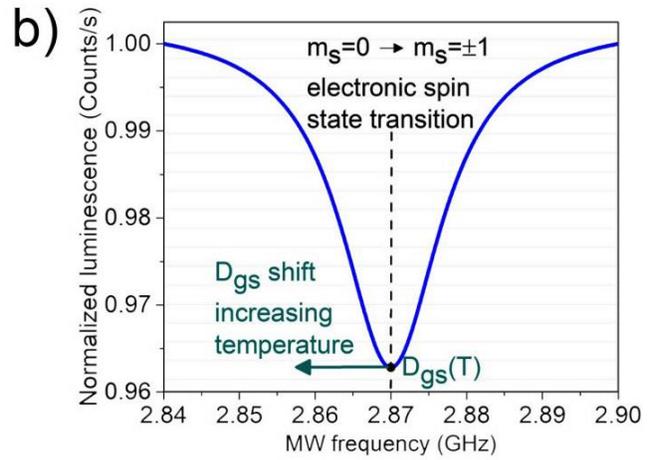
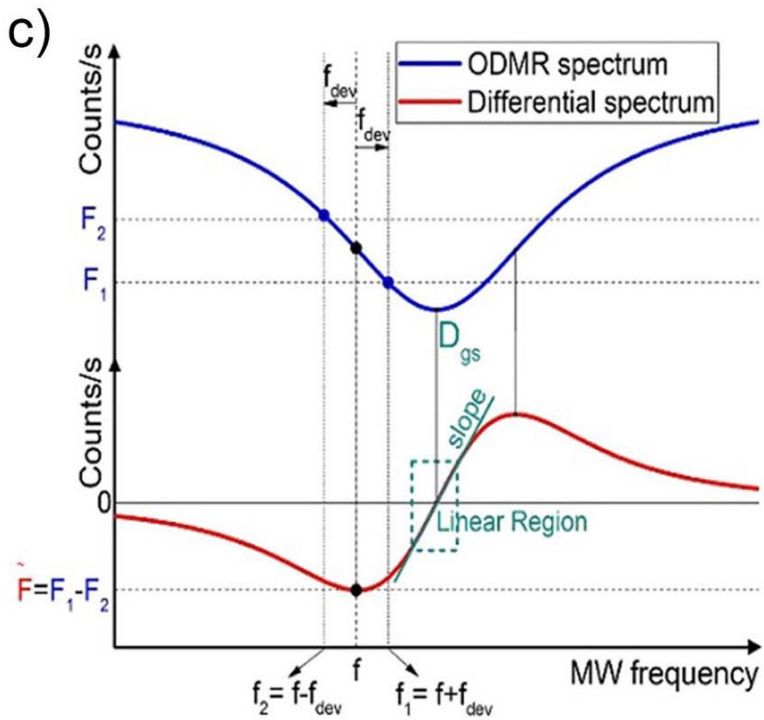
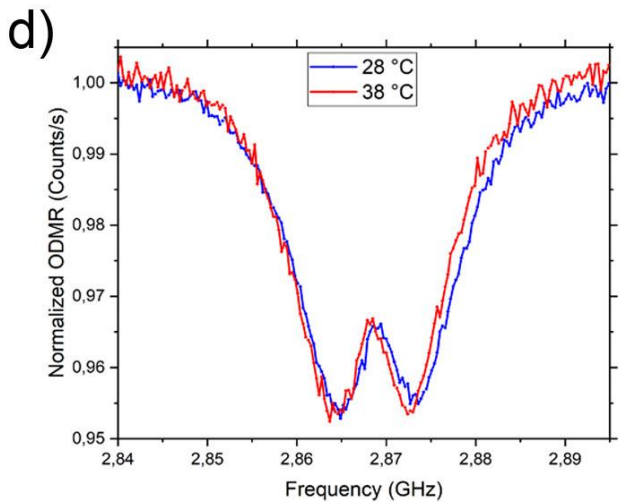
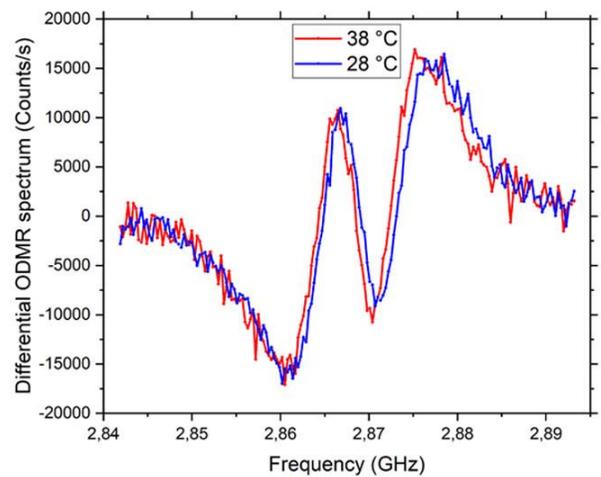

**Figure 4** a) NV⁻ state transition that occurs after laser excitation and MW excitation. The coupling of the states |m$_s$ =±1>; with the metastable level generates a statistically lower PL emission than the |m$_s$ = 0 >. b) PL collected from the NV⁻ center as a function of the MW frequency. A dip in correspondence of the zero-field splitting D$_{gs}$ (resonance frequency of the undisturbed NV⁻ center, at room temperature) can be observed. c) Sketch of the differential measurement. From the ODMR spectrum (upper part of the figure) the differential spectrum (lower part of the figure) is derived taking, for every value of the microwave (MW) frequency, the difference in PL $\tilde{F}$ between two points separated by 2f$_{dev}$ . $\tilde{F}$ is zero at the two extremes of the spectrum and at the resonant frequency D$_{gs}$. Around D$_{gs}$ there is a region where $\Delta\tilde{F}$ depends linearly on $\Delta D_{gs}$ through the differential spectrum slope. d) Example of ODMR and differential ODMR spectra at two different temperatures.

## 4.6 Estimation of $\partial D_{gs}/\partial T$

Under ambient conditions the coupling constant $\partial D_{gs}/\partial T$ is estimated as -75 kHz/°C for a bulk diamond Under ambient conditions the coupling constant $\partial D_{gs}/\partial T$ is estimated as -75 kHz/°C for a bulk diamond[42,43]. For NDs this constant varies due to the different strain splitting and, in principle, should be calibrated.

The method to obtain $\partial D_{gs}/\partial T$ was the same adopted for the cell temperature measurement, with the difference that in this case $\Delta T$ was not induced by a substance perfusion but forced by the incubator chamber's heater, and that $\Delta T$ is measured by a thermocouple. The procedure was the following. The NDs were placed on the Petri dish with distilled water. The ODMR spectrum was acquired and so was the differential spectrum by post-processing the data, see above. The linear region *slope* of the differential spectrum was obtained by performing a linear regression. The microwave generator was set at the resonant frequency $D_{gs,T_0}$ that occured in the initial temperature condition reached in the incubator chamber, at which the differential spectrum signal $\tilde{F}$ was zero. The incubator temperature was varied, recording it with the thermocouple. The new value of the differential signal was read. Finally, $\Delta\tilde{F}$ was plotted as a function of the temperature increase $\Delta T$ recorded by the thermocouple. Through a linear regression the constant of proportionality $b$ was evaluated, i.e. the quantities mentioned obey the following physical relationship:

$$\Delta\tilde{F} = b \cdot \Delta T \tag{4}$$

where:

$$b = slope \cdot \partial D_{gs}/\partial T \qquad (5)$$

By inverting the last equation the quantity $\partial D_{gs}/\partial T$ was estimated. The results are shown in Table S1. Both the arithmetic mean and the weighted mean coincide: $\partial D_{gs}/\partial T = (-76 \pm 4) \, kHz/°C$. We used this estimated value for $\partial D_{gs}/\partial T$ and the associated uncertainty in our data analysis.

### 4.7 Cell preparation and ND labeling

Hippocampal neurons were obtained from WT 16-day embryos. Hippocampus was rapidly dissected under sterile conditions, kept in cold HBSS (4°C) with high glucose, and then digested with papain (0,5 mg/ml) dissolved in HBSS plus DNAse (0.1 mg/ml). Isolated cells were plated at density of 1200 cells/mm$^2$ onto the MEA and 1000 cells/mm$^2$ onto the glass Petry dishes. Both the MEAs and the dishes were previously coated with poly-DL-lysine and laminin, this allowed the neurons to adhere to the center of the device by using a ring made of Sylgard 184 (Dow Corning), which was removed after 4 h. The cell medium is composed of 1% penicillin/streptomycin, 1% glutamax, 2.5% fetal bovine serum, 2% B-27 supplemented neurobasal medium. The neurons were incubated in a humidified 5% $CO_2$ atmosphere at 37°C. Recordings were carried out at 10-12 days in Vitro (DIV).

The initial ND solution (1 mg/ml, MiliQ water) was sonicated for 5-10 minutes (100 W power, 80 kHz frequency), in order to separate diamond particle clusters into single NDs. After sonication, the 60 μl of solution was diluted with 1 ml of Tyroide to obtain a final ND concentration of 0.6 μg/ml which was poured into the Petri dish. The dish was exposed to NDs for 5 hours in order to allow NDs internalization. Once the dish was extracted from the incubator, the cell medium was removed and replaced with 2 ml of Tyrode solution.

### 4.8 MEA Recordings

Multisite extracellular recordings were performed using the MEA-system, purchased from Multi-Channel Systems (Reutlingen, Germany). The 60 electrodes array (TiN) was composed of an 8 × 8 square grid with 200 μm inter-electrode spacing and 30 μm electrode diameter. Data acquisition was controlled through MC_Rack software (Multi-Channel Systems Reutlingen, Germany), sampling at 10 kHz. Experiments were performed in a non-humidified incubator at 37°C and with 5% $CO_2$, maintaining the culture medium. Before starting the experiments, cells were allowed to stabilize in the non-humidified incubator for 5 minutes; recording of the spontaneous activity was then carried out for 120 s. Mean frequency has been evaluated over 120 sec recording. The data are analyzed using Neuroexplorer software (Nex Technologies, Littleton, MA, USA) after spike sorting operations.

## 4.9 Confocal image acquisition

After 5 hours exposure to ND (6 μg/ml NDs), the cytoplasmic labelling dye (CellTracker™ Green CMFDA, ThermoFisher,) was added to the medium. This allowed to identify the cell boundaries of cultured hippocampal neurons (10 DIV) together with the internalized NDs, characterized by a red fluorescent emission. For standard confocal imaging, hippocampal neurons were plated on 35 mm dishes (ibidi GmbH, Planegg/ Martinsried, Germany). Cells were analyzed using a confocal laser scanning microscope Leica TCS SP5, equipped with an argon ion and a 561 nm DPSS laser. Cells were imaged using a HCX PL APO 63x/1.4 NA oil immersion objective at a pixel resolution of 0.08 x 0.08 x 0.3 μm. The luminescent emission from the FNDs was excited by 561 nm laser, while the emission was collected in the 650–750 nm spectral range. The same excitation wavelength was ineffective in untreated neurons. Green fluorescence for intracellular staining was obtained by 488 nm wavelength. Image analysis was performed using ImageJ software.

## 4.10 Nanodiamonds preparation

*4.10.1 Monodisperse ND*

NDs were supplied by Microdiamant Switzerland (MSY 0–0.25, containing approximately 100–200 ppm of natural nitrogen impurities) and treated as described previously *(40)*. Briefly, the NDs were oxidized by air oxygen at 510°C for 5 h and then wet oxidized in a HF:HNO$_3$ 2:1 v/v stirred mixture at 160°C for 2 days in a PTFE container. The acids were removed using consecutive centrifugation/washing and the resulting pure oxidized NDs were lyophilized. The monodisperse NDs (hydrodynamic diameter 205 nm) were isolated using differential centrifugation in water based on gradual centrifugation/dilution steps, combination of the separated pellets, followed by their lyophilization.

*4.10.2 Fluorescent NDs: irradiation and oxidation*

A total of 330 mg of the monodisperse ND was irradiated at 870 °C [45] in an external target for 80 h with a 15.7 MeV electron beam (2.5·10$^{19}$ particles cm$^{-2}$) extracted from the MT-25 microtron. After irradiation, the NDs (315 mg) were annealed at 900°C for 1 h under argon atmosphere and subsequently oxidized in air for 5 h at 510°C. The resulting FNDs were wet oxidized using a 2:1 (v/v) mixture of HF:HNO$_3$, washed and lyophilized, providing 212 mg of NDs with NV centers (yield 64 %). Before use, the lyophilized NDs were redispersed in MiliQ water using cup horn sonication to concentration 1.0 mg ml$^{-1}$.

*4.10.3 Dynamic light scattering*

The hydrodynamic particle diameter was measured using a Zetasizer Nano ZS (Malvern Instruments) in disposable transparent cuvettes. The samples were measured at RT with a concentration of 0.05 mg/ml and a total volume of 1 ml.

**4.11 Nanodiamond sensor sensitivity**

In order to evaluate the sensitivity of our ND-based sensors, a control calibration test was performed. The NDs were deposited on a glass slide and inserted into the temperature chamber of the single-photon confocal microscope. The differential signal was evaluated at three different temperatures and then compared with the readout of a standard thermocouple. Figure 5 shows the response of the ND sensor to a thermal cycle with steps 0.5°C, 1.2°C. The average values are shown in the figure next to each dataset. The results are in agreement with the bath temperature recorded by the thermocouple, shown in the legend box. This control test highlights the sensitivity of the measurement method, capable of discriminating a variation of 0.5°C. Our sensitivity is about $3 \frac{°C}{\sqrt{Hz}}$ and is estimated according to the following equation:

$$\eta = \frac{\sigma_{\tilde{F}} \sqrt{\Delta t}}{slope \; \gamma_T} \tag{6}$$

where $\sigma_{\tilde{F}}$ represents the standard error of the mean of the variation of the differential counts and $\Delta t$ represents the measurement time.

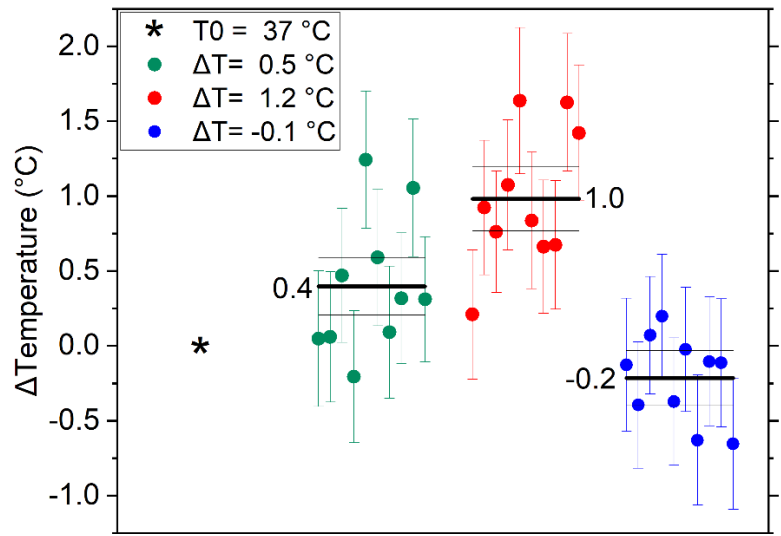

**Figure 5** Validation of temperature detection by ND sensor. The legend box shows the temperature values recorded by the thermocouple. The box plot shows the temperature values recorded by the ND sensor. The mean and the uncertainty are reported as horizontal lines in each data set. The starting point is highlighted as an asterisk. The incubator temperature at the end of the cycle is consistent with the initial temperature within the incubator stability.

## 4.12 Statistic uncertainty analysis

As detailed in the main text, the experiment was repeated collecting a sample of N=11 repeated measurements, revealing any temperature rise from a ND inside the cell before and after tyrode perfusion. In the main test this is referred to as "CTRL". In another sample of N=10 NDs the temperature was probed by a ND inside the cell before and after perfusing picrotoxin. This sample is referred to as "+PICRO". Starting from this condition, a sample of N=8 temperature variation repeated measurements were performed in a ND inside the cell before and after TTX and cadmium perfusion. This is referred to as "+PICRO+TTX+Cd".

The temperature variation ΔT detected by the NV sensor was evaluated through a direct fluorescence change measurement, as described before, the equation being:

$$\Delta T = \frac{\Delta \tilde{F}}{slope \cdot \partial D_{gs}/\partial T} \tag{7}$$

Every single measure of temperature change $\Delta T_i$ was affected by an uncertainty $\sigma(\Delta T_i)$. For its estimation, the error propagation of the variables that appear in Equation 7 has been carried out. In the following, the involved uncertainties are analyzed in detail.

The term $\Delta \tilde{F}$ is the difference between the differential spectrum signal $\tilde{F}$ before and after the substance perfusion. The associated uncertainty is the sum in quadrature: $\sigma(\Delta \tilde{F}) = \sqrt{[\sigma(\tilde{F}_{after})]^2 + [\sigma(\tilde{F}_{before})]^2}$.

Where $\sigma(\tilde{F})$ is data standard deviation divided by the square root of repeated measurements (500 values in 60 s).

As well as the PL, the slope was also evaluated on every single measurement. The associated uncertainty $\sigma(slope)$ was evaluated by the linear regression of the linear region differential spectrum.

The constant $\partial D_{gs}/\partial T$ was evaluated by a previous measurement, analyzing a sample of N=6 NDs (Table 1) as described in the Section "Method to obtain $\partial D_{gs}/\partial T$". Unlike previous contributions, its associated uncertainty acts as a *b type* uncertainty, intervening only in the errors propagation on the average value estimate of each data set. The values $\Delta T_i$ and $\sigma(\Delta T_i)$ are shown in Table 2.

To prove the main result of this work it was necessary to demonstrate a significant statistical difference between the "+PICRO" and the "CTRL" sets. This can be formalized by evaluating the probability that

$\Delta T_{+PICRO}$ and $\Delta T_{CTRL}$ belong to the same population, which is the starting null hypothesis. The *Welch t-student* test was performed for this purpose. The calculated values for the stochastic *t* variable, the population degree of freedom and the probability of satisfying the null hypothesis are shown in Table 3.

The test significance obtained allowed rejecting the initial null hypothesis. The two samples did not belong to the same population and therefore the increase in temperature, recorded following picrotoxin perfusion, was statistically significant. The same test was repeated to compare the "CTRL" and "+PICRO+TTX+Cd" sets in order to understand if the decrease in temperature recorded by the nanodiamonds following the TTX and Cd perfusion is statistically significant. As can be seen from Table 3, also in this case it is possible to reject the null hypothesis. The two samples did not belong to the same population and therefore the decrease in temperature was statistically significant.

| $\frac{\partial D_{gs}}{\partial T}\left[\frac{kHz}{°C}\right]$ |
| --- |
| 79 ± 6 |
| 63 ± 9 |
| 76 ± 12 |
| 86 ± 12 |
| 72 ± 8 |
| 81 ± 10 |

**TABLE 1.** Summary of NDs (№=6) coupling constant evaluations and their associated uncertainty, obtained through the uncertainty propagation

| $\Delta T \pm \sigma(\Delta T)$ [°C] | | |
|---|---|---|
| CTRL | +PICRO | +PICRO+TTX+Cd |
| -0.08 ± 0.77 | 1.39 ± 0.76 | 0.18 ± 0.80 |
| -0.27 ± 0.49 | 1.38 ± 1.02 | 0.14 ± 0.48 |
| -0.17 ± 0.47 | 1.01 ± 0.90 | -0.63 ± 0.29 |
| -0.60 ± 1.23 | 0.65 ± 0.64 | -1.46 ± 0.72 |
| 0.33 ± 0.33 | 1.73 ± 0.85 | -0.76 ± 0.60 |
| 0.42 ± 0.88 | 1.04 ± 0.65 | -0.92 ± 0.57 |
| 0.18 ± 0.59 | 1.12 ± 1.10 | -0.20 ± 0.31 |
| -0.07 ± 0.49 | 0.70 ± 1.06 | -1.50 ± 0.77 |
| 0.29 ± 0.33 | 1.04 ± 0.49 | |
| -0.47 ± 0.55 | 0.73 ± 0.57 | |
| -0.01 ± 0.44 | | |

**TABLE 2.** Summary of temperature variation $\Delta T \pm \sigma \Delta T$ [°C] for the three independent groups.

| CTRL vs +PICRO | |
|---|---|
| Welch's t value | 11.24 |
| Degree of freedom | 14.90 |
| Significance Welch's t-test | $1.12 \cdot 10^{-8}$ |

| CTRL vs +PICRO+TTX+Cd | |
|---|---|
| Welch's t value | 7.63 |
| Degree of freedom | 14.15 |
| Significance Welch's t-test | $2.21 \cdot 10^{-6}$ |

**TABLE 3.** Summary of statistical parameters for the Welch's t-test.

**Acknowledge:** We acknowledge Prof. E. Carbone for useful discussion and C. Franchino for cell preparation This project (20IND05) QADeT leading to this publication has received funding from the EMPIR program co-financed by the Participating States and from the European Union's Horizon 2020 research and innovation program. European Commission's PATHOS EU H2020 FET-OPEN grant no. 828946 Horizon 2020, MSM Project No. 8C18004 (NanoSpin), ChemBioDrug (No. CZ.02.1.01/0.0/0.0/16_019/0000729), CARAT (No.CZ.02.1.01/0.0/0.0/16_026/0008382);

**Authors contributions:** MG, VC, EB, EM, IPD, PC and PT planned the experiment. The main experimental work was equally realized by GT (for the biological part) and GP (optical one) with the direct help and supervision of EB and EM (for the optical one) in the laboratories directed by VC (biological one) and MG (optical one). The quantum sensing facility at INRIM was developed by EM, EB and PT. All the authors equally contributed to the discussion on experimental settings and on data analysis, as well as in writing the paper. PC and KK provided the nano-diamond samples;

**Competing interests:** Authors declare that they have no competing interests.

**Data and materials availability:** The data that support the findings of this study are available from the corresponding author on reasonable request.

## References

[1]     B. Hille, *Ion Channels of Excitable Membranes*, Sinauer Associates, Inc., Sunderland, Massachusetts, **2001**.

[2]     T. O'Leary, E. Marder, *Curr. Biol.* **2016**, *26*, 2935.

[3]     L. Bui, M. I. Glavinović, *Cogn. Neurodyn.* **2014**, *8*, 277.

[4]     D. G. Margineanu, E. Schoffeniels, *Proc. Natl. Acad. Sci. U. S. A.* **1977**, *74*, 3810.

[5]     J. V. Howarth, J. M. Ritchie, D. Stagg, *Proc. R. Soc. London - Biol. Sci.* **1979**, *205*, 347.

[6]     A. C. L. De Lichtervelde, J. P. De Souza, M. Z. Bazant, *Phys. Rev. E* **2020**, *101*, 22406.

[7]     J. Choi, H. Zhou, R. Landig, H. Y. Wu, X. Yu, S. E. von Stetina, G. Kucsko, S. E. Mango, D. J. Needleman, A. D. T. Samuel, P. C. Maurer, H. Park, M. D. Lukin, *Proc. Natl. Acad. Sci. U. S. A.* **2020**, *117*, 14636.

[8]     L. De Meis, L. A. Ketzer, R. M. Da Costa, I. R. De Andrade, M. Benchimol, *PLoS One* **2010**, *5*, DOI 10.1371/journal.pone.0009439.

[9]     S. M. Thompson, L. M. Masukawa, D. A. Prince, *J. Neurosci.* **1985**, *5*, 817.

[10]    M. Volgushev, T. R. Vidyasagar, M. Chistiakova, U. T. Eysel, *Neuroscience* **2000**, *98*, 9.

[11]    J. C. F. Lee, J. C. Callaway, R. C. Foehring, *J. Neurophysiol.* **2005**, *93*, 2012.

[12]    M. E. Raichle, *Proc. Natl. Acad. Sci. U. S. A.* **1998**, *95*, 765.

[13]  M. Monti, L. Brandt, J. Ikomi-Kumm, H. Olsson, *Scand. J. Haematol.* **1986**, *36*, 353.

[14]  M. Ghavami, M. Rezaei, R. Ejtehadi, M. Lotfi, M. A. Shokrgozar, B. Abd Emamy, J. Raush, M. Mahmoudi, *ACS Chem. Neurosci.* **2013**, *4*, 375.

[15]  P. Liang, Y. Xu, X. Zhang, C. Ding, R. Huang, Z. Zhang, J. Lv, X. Xie, Y. Chen, Y. Li, Y. Sun, Y. Bai, Z. Songyang, W. Ma, C. Zhou, J. Huang, *Protein Cell* **2015**, *6*, 363.

[16]  M. Fujiwara, S. Sun, A. Dohms, Y. Nishimura, K. Suto, Y. Takezawa, K. Oshimi, L. Zhao, N. Sadzak, Y. Umehara, Y. Teki, N. Komatsu, O. Benson, Y. Shikano, E. Kage-Nakadai, *Sci. Adv.* **2020**, *6*, 1.

[17]  J.-M. Yang, H. Yang, L. Lin, *ACS Nano* **2011**, *5*, 5067.

[18]  S. Kiyonaka, T. Kajimoto, R. Sakaguchi, D. Shinmi, M. Omatsu-Kanbe, H. Matsuura, H. Imamura, T. Yoshizaki, I. Hamachi, T. Morii, Y. Mori, *Nat. Methods* **2013**, *10*, 1232.

[19]  D. Chrétien, P. Bénit, H. H. Ha, S. Keipert, R. El-Khoury, Y. T. Chang, M. Jastroch, H. T. Jacobs, P. Rustin, M. Rak, *PLoS Biol.* **2018**, *16*, 1.

[20]  G. Baffou, H. Rigneault, D. Marguet, L. Jullien, *Nat. Methods* **2014**, *11*, 899.

[21]  M. Suzuki, V. Zeeb, S. Arai, K. Oyama, S. Ishiwata, *Nat. Methods* **2015**, DOI 10.1038/nmeth.3551.

[22]  M. Suzuki, T. Plakhotnik, *Biophys. Rev.* **2020**, *12*, 593.

[23]  F. Vetrone, R. Naccache, A. Zamarrón, A. J. De La Fuente, F. Sanz-Rodríguez, L. M. Maestro, E. M. Rodriguez, D. Jaque, J. G. Sole, J. A. Capobianco, *ACS Nano* **2010**, *4*, 3254.

[24]  M. Li, T. Chen, J. J. Gooding, J. Liu, *ACS Sensors* **2019**, *4*, 1732.

[25]  S. W. Allison, G. T. Gillies, A. J. Rondinone, M. R. Cates, *Nanotechnology* **2003**, *14*, 859.

[26] L. Guarina, C. Calorio, D. Gavello, E. Moreva, P. Traina, A. Battiato, S. Ditalia Tchernij, J. Forneris, M. Gai, F. Picollo, P. Olivero, M. Genovese, E. Carbone, A. Marcantoni, V. Carabelli, *Sci. Rep.* **2018**, *8*, 2221.

[27] G. Petrini, E. Moreva, E. Bernardi, P. Traina, G. Tomagra, V. Carabelli, I. Pietro Degiovanni, M. Genovese, *Adv. Quantum Technol.* **2020**, DOI 10.1002/qute.202000066.

[28] T. Zhang, G. Pramanik, K. Zhang, M. Gulka, L. Wang, J. Jing, F. Xu, Z. Li, Q. Wei, P. Cigler, Z. Chu, *ACS Sensors* **2021**, *6*, 2077.

[29] M. Fujiwara, Y. Shikano, *Nanotechnology* **2021**, *32*, 482002.

[30] C. Bradac, S. F. Lim, H. Chang, I. Aharonovich, *Adv. Opt. Mater.* **2020**, *8*, DOI 10.1002/adom.202000183.

[31] I. V. Fedotov, M. A. Solotenkov, M. S. Pochechuev, O. I. Ivashkina, S. Y. Kilin, K. V. Anokhin, A. M. Zheltikov, *ACS Photonics* **2020**, DOI 10.1021/acsphotonics.0c00706.

[32] Y. Wu, T. Weil, *Adv. Sci.* **2022**, 220005.

[33] G. Kucsko, P. C. Maurer, N. Y. Yao, M. Kubo, H. J. Noh, P. K. Lo, H. Park, M. D. Lukin, *Nature* **2013**, *500*, 54.

[34] D. A. Simpson, E. Morrisroe, J. M. McCoey, A. H. Lombard, D. C. Mendis, F. Treussart, L. T. Hall, S. Petrou, L. C. L. Hollenberg, *ACS Nano* **2017**, *11*, 12077.

[35] L. Guarina, C. Calorio, D. Gavello, E. Moreva, P. Traina, A. Battiato, S. Ditalia Tchernij, J. Forneris, M. Gai, F. Picollo, P. Olivero, M. Genovese, E. Carbone, A. Marcantoni, V. Carabelli, *Sci. Rep.* **2018**, DOI 10.1038/s41598-018-20528-5.

[36] V. Tseeb, M. Suzuki, K. Oyama, K. Iwai, S. Ishiwata, *HFSP J.* **2009**, DOI 10.2976/1.3073779.

[37] A. Marcantoni, M. S. Cerullo, P. Buxeda, G. Tomagra, M. Giustetto, G. Chiantia, V. Carabelli, E. Carbone, *J. Physiol.* **2020**, *598*, 2183.

[38] D. Attwell, S. B. Laughlin, *J. Cereb. Blood Flow Metab.* **2001**, 1133.

[39] B. E. Leonard, *Hum. Psychopharmacol. Clin. Exp.* **1993**, DOI 10.1002/hup.470080412.

[40] D. Gatto Monticone, K. Katamadze, P. Traina, E. Moreva, J. Forneris, I. Ruo-Berchera, P. Olivero, I. P. Degiovanni, G. Brida, M. Genovese, *Phys. Rev. Lett.* **2014**, *113*, DOI 10.1103/PhysRevLett.113.143602.

[41] S. M. Smith, *Hum. Brain Mapp.* **2002**, *17*, 143.

[42] X. D. Chen, C. H. Dong, F. W. Sun, C. L. Zou, J. M. Cui, Z. F. Han, G. C. Guo, *Appl. Phys. Lett.* **2011**, *99*, 161903.

[43] V. M. Acosta, A. Jarmola, E. Bauch, D. Budker, *Phys. Rev. B - Condens. Matter Mater. Phys.* **2010**, DOI 10.1103/PhysRevB.82.201202.

[44] V. M. Acosta, E. Bauch, M. P. Ledbetter, A. Waxman, L. S. Bouchard, D. Budker, *Phys. Rev. Lett.* **2010**, *104*, 070801.

[45] M. Capelli, A. H. Heffernan, T. Ohshima, H. Abe, J. Jeske, A. Hope, A. D. Greentree, P. Reineck, B. C. Gibson, *Carbon N. Y.* **2019**, *143*, 714.